 \documentclass[nohyper,12pt,letterpaper]{JHEP3}
 \usepackage{epsfig}
 
 \def\bR{{\mathbb R}}
 \def\bZ{{\mathbb Z}}

 \title{Populating the swampland: \\the case  of  $U(1)^{496}$ and $E_8\times U(1)^{248}$}
 \author{Bartomeu\,Fiol\\
 Departament de F{\'\i}sica Fonamental i \\Institut de Ci{\`e}ncies del Cosmos, 

Universitat de Barcelona,

Mart{\'\i}\ i Franqu{\`e}s 1, 08193 Barcelona, Catalonia, Spain.\\

\email{bfiol@ub.edu}}

\abstract{For $d=10$ ${\cal N}=1$ SUGRA coupled to $d=10$ ${\cal N}=1$ SYM, anomaly 
cancellation places severe constraints on the allowed gauge groups. Besides the ones
known to appear in string theory, only $U(1)^{496}$ and $E_8\times U(1)^{248}$ are 
allowed. There are no known theories of quantum gravity that reduce in some limit to these two last supergravity theories, and in this note I present some evidence that those quantum theories might not exist. The first observation is that, upon compactification, requring that the quantum theory possesses a moduli space with finite volume typically implies the existence of singularities where the 4d gauge group is enhanced, but for these two theories that gauge enhancement is problematic from the 10d point of view.  I also  point out that while these four supergravity theories present repulson-type singularities, the known mechanism that repairs those singularities for the first two - the non-Abelian enhan\c{c}on - is not available for the last two theories. In short, these two supergravity theories might be too Abelian for their own good. }
 
\begin{document}
\section{\bf Introduction and conclusions}

The swampland program aims at delimiting the boundary that separates the  effective theories of gravity that admit a full embedding in a theory of quantum gravity from those that don't \cite{Vafa:2005ui,   Ooguri:2006in}. If we don't assume that M/string theory is {\it the} quantum theory of gravity, the toolkit at our disposal to carry out this program seems quite limited. The best such established tool is anomaly cancellation:  even though anomalies are a quantum effect, there are necessary conditions than any classical theory must satisfy in order to be anomaly free, and these conditions can be written in terms of just the classical (massless) field content of the theory.

If the swampland program is to make any progress, it seems urgent to extend the toolkit used to diagnose if a gravity theory belongs to the swampland or not. Over the years, a variety of pathologies have been found for classical theories of gravity: closed timelike curves, naked singularities, and so on.
Their respective fate in quantum gravity is far from settled, although there are many interesting suggestions \cite{Hawking:1991nk, Gubser:2000nd, Natsuume:2001ba, Bojowald:2007ky}. For instance, it has been argued \cite{Horowitz:1995ta} that not every naked singularity in classical gravity should be resolved in the quantum theory. My working philosophy is to hope that there do exist classical pathologies that {\it must} be resolved in the quantum theory for it to be consistent, and one should try to pin them down.

In order to do field work, and look for possible patterns in the boundary between the landscape and the swampland, a good idea seems to start with as much supersymmetry as possible. In \cite{Green:2007zzb} it was argued that 4d ${\cal N}=8$ SUGRA belongs to the swampland. If we now look at SUGRA theories with 16 supercharges, the possibilities in 10d are very few. For 10d $ {\cal N}=1$ SUGRA coupled to ${\cal N}=1$ SYM, anomaly cancellation restricts the possible SYM group to $SO(32)$, $E_8\times E_8$, $U(1)^{496}$ and $E_8\times U(1)^{248}$ \cite{gsw}. There is no known string theory  -or for that matter, any other quantum theory of gravity - that reduces to the last two supergravities. Is this just because we didn't look hard enough, or can we come up with some reason why such quantum theories can't exist?

This work doesn't provide a rigorous answer to the preceeding question, but it points out some differences between the two pairs of supergravity theories, that suggest that the last two belong to the
swampland. One feature of the last two theories that immediately sticks out, compared with the heterotic ones, is that their gauge group contains a large Abelian factor. To stress this fact, in what follows I will often refer to the first pair as the heterotic theories, and to the last two as the Abelian ones (even though $U(1)^{248}\times E_8$ is of course non-Abelian). Since we have at our disposal quantum theories for the first pair of supergravity theories, the strategy seems clear: we should look for features/pathologies common to these four supergravity theories, that for the first pair are resolved in the quantum theory thanks to the non-Abelian character of their 10d gauge group. Then, these pathologies aren't readily resolved in the Abelian theories and, if we grant these pathologies ought to be absent in quantum gravity, it strengthens the suspicion that the Abelian theories belong to the swampland. These two supergravity theories might be 'too Abelian' for their own good.

A first observation is that these four supergravity theories have a classical moduli space with infinite volume. On the other hand, it has been argued in various contexts \cite{Horne:1994mi, Vafa:2005ui, Douglas:2005hq} that theories of quantum gravity ought to have moduli spaces with finite volume.
For the first two pair of supergravity theories, it is well known how this comes about at the quantum level: the classical moduli space gets quotiented by a lattice, and the resulting fundamental domain has finite volume. The process of quotienting introduces various singularities in moduli space, and the physics of those singularities is understood: states that are generically massive become massless at those points in moduli space; in particular some such singularities correspond to $W^\pm$ boson multiplets of the 10d gauge group, becoming massless and enhancing the 4d gauge group. Clearly that type of singularity would be problematic in a hypothetical quantum moduli space of one of the Abelian theories, as now compactification of the gauge group doesn't produce massive $W^\pm$ bosons ready to go massless at the singularity, and the 10d theory can't afford any more gauginos, due to anomaly cancellation. If one were able to argue that {\it any} physically reasonable quotient of the classical moduli spaces that yields finite volume, produces such singularities, that would be a strong strike against the Abelian theories. In the next section, I outline the rudiments of such an argument.

The next observation I will discuss is in a sense related to the previous one, but shifting gears from singularities in moduli space to singularities in spacetime solutions. The toroidal compactifications of these four $10d$ SUGRA theories present solutions with naked singularities, the so-called repulson singularities \cite{Behrndt:1995nn, Kallosh:1995yz, Behrndt:1995nn, Cvetic:1995mx}. Those solutions are charged under some $U(1)$, which four our purposes we take to be in the Cartan subalgebra of the 10d gauge group, and for the Abelian theories, in the Abelian factor of the gauge group. For the first two SUGRA theories, there is a well-known mechanism to repair these naked singularities, the enhan\c {c}on mechanism \cite{Johnson:1999qt}, but as we will argue, this mechanism, being non-Abelian,  does not readily apply to the Abelian SUGRA theories. 

The two Abelian supergravities have solutions with naked singularities that can't be repaired by the mechanism that works for the known quantum realization, namely string theory, of the other two
supergravities. What should we conclude from this observation? After all, there are singularities in classical gravity that are expected not to be repaired in quantum gravity, like the negative mass Schwarzschild black hole \cite{Horowitz:1995ta}. A first difference is that the repulson singularity has healthy-looking asymptotic charges, i.e. it's like an overcritical Reissner-Nordstrom black hole; we are used to configurations with such charge/mass relation, we call them electrons. Having a sugra solution with healthy-looking charges is not enough guarantee that there are states in the quantum theory with those charges. At this point, we appeal to supersymmetry; the solutions we will be considering are $1/2$ BPS solutions in a theory with 16 supercharges, and although they can get corrected, they are expected to persist at the quantum level. Finally, at least in the cases where we do have a quantum theory that in the supergravity limit presents such naked singularities, we know for a fact that the quantum theory repairs them. All in all, we take these bits as circumstancial evidence that these extremal repulson singularities ought to be resolved in quantum gravity.

An obvious possibility to consider is that the putative quantum theory for the Abelian sugras does repair the repulson singularities by another mechanism different from the enhan\c {c}on. Of course, I can't rule this out completely. I will just recall below that a purely supergravity analysis \cite{Johnson:2001wm} of the repulson solution puts constraints on the kinds of excisions of the singularity one might consider, and the properties of the stress-energy tensor at the junction, if one does not want to violate any energy condition. Finally, it is possible that these naked singularities can't be dynamically formed in the Abelian theories, starting from non-singular configurations, and that in quantum gravity, this kind of eternal naked singularities are allowed.

\section{Finiteness of the quantum moduli spaces.}
One of the conjectures put forward in the swampland program is that moduli spaces that arise upon honest compactifications (i.e. those that give a non-zero gravitational constant in the lower dimension) have finite volume.  This requirement was discussed in \cite{Horne:1994mi} in the context of fine tuning of the initial conditions in string cosmology, and more recently has resurfaced in studies of the string landscape, since the volume of moduli space appears in the estimates of the number of flux vacua  \cite{Ashok:2003gk}. Finite volume  is a non-trivial requirement, since these moduli spaces classically can have infinite volume, so it demands the existence of quantum symmetries that identify different points of the classical moduli space. 

For the first two theories, we have quantum realizations (heterotic strings and their type II duals), for which those symmetries are well-known, T-duality, (and in 4d, S-duality).  In what follows I will present an argument suggesting that for the second two theories, having a quantum realization with finite volume moduli space might be quite a tall order \footnote{It is possible that there exist lattices of $SO(d,d; \bR)$ acting freely on the classical moduli space, i.e. providing a moduli space with finite volume and without singularities. If they exist, it would be important to understand what kind of spectrum they correspond to, and if they can be discarded on physical grounds, e.g. by checking if they satisfy the other conjectures of \cite{Ooguri:2006in}.}. Since I am not assuming that the quantum theory is a string theory, I won't rely on any world-sheet argument. 

The four SUGRA theories at hand, after toroidal compactification on $T^d$ have many massless scalar fields, parameterized (aside from the dilaton, and in 4d, the axion) by classical moduli spaces of the form
\begin{equation}
\frac{SO(d,d+r;\bR)}{SO(d;\bR)\times SO(d+r;\bR)}
\label{modul}
\end{equation}
These moduli spaces are Riemannian manifolds with constant negative curvature, and have infinite volume. I will assume that in the putative quantum theory, the moduli space is still locally of the form 
(\ref{modul}), so it is quotiented by a discrete group $\Gamma$ of $SO(d,d+r;\bR)$. For generic discrete subgroups, the volume of the quotient will still be infinite, but if $G$ is a non-compact simple group and $K$ its maximal compact subgroup,  $\Gamma \backslash G/K$ has finite volume if and only if $\Gamma$ is a lattice in $G$ (i.e $\Gamma \backslash G$ has finite volume). So I take the requirement of finite volume as to imply that in the quantum theory, the classical moduli space (\ref{modul}) gets quotiented by a lattice (and not just any discrete group) $\Gamma$ of $SO(d,d+r;\bR) $.

Besides the scalar fields, at generic points in moduli space, these compactified sugra theories have gauge group is $U(1)^{2d+r}$. Classically, a rotation in the moduli matrix is accompanied by a rotation in the gauge fields
$$
A_\mu \rightarrow \Omega A_\mu   \hspace{1cm} \Omega \in SO(d,d+r)
$$
Physically, we take the restriction to $\Gamma$ at the quantum level as signaling the existence of massive particles charged under $U(1)^{2d+r}$,  their quantized charges living in $\Gamma$. What this doesn't tell us is which sites of the lattice actually correspond to states in the theory, and to what multiplets these states might belong. We want to further argue that among them, there are $\frac{1}{2}$ BPS ${\cal N}=4$ vectormultiplets (i.e. $W^+$ gauge bosons and their partners).

Let's assume that there is a putative quantum theory of gravity reducing at low energies to either of the two Abelian supergravity theories. We know the part of the spectrum that is massless everywhere in moduli space, but we don't know its massive spectrum. Even for BPS states, the mass formula derived in SUGRA tells us that if a state exists, its mass is determined, but it does not tell us which states actually exist in the spectrum of the theory. 

When compactifying the supergravity from ten to lower dimensions, we get towers of $1/2$ BPS
KK states, and I will take for granted that these states remain in the quantum theory. They are electrically charged, with a quantized charge, with respect to the KK $U(1)$s . At least for large momentum charge, their mass can be computed in SUGRA, and due to nonrenormalization arguments,  it is not expected to
get quantum corrections
$$
m^2_{KK}=\left(\frac{n_1}{R_1}\right)^2 +\dots + \left(\frac{n_d}{R_d}\right)^2 
$$
To have a finite volume moduli space in the quantum theory, we need to identify points in the space of radii $R_i$. An obvious symmetry is permutation exchange between radii $R_i \leftrightarrow R_j$, but this is not enough to give finite volume. An important feature of the identifications we are after is that, at the points identified,  the full spectrum (massless and massive) of the theory coincides. If the KK spectrum is the only massive spectrum of the theory, there is no identification of points in the space of radii that leaves the spectrum invariant. So demanding that the quantum moduli spaces have finite volume seems to imply that KK states can't be alone in the massive spectrum. Looking at the BPS mass formula, a solution presents itself: the existence of 1/2 BPS massive states charged under $B_\mu$. They have mass $wR$ and will allow for a compact moduli space if we identify $R\sim 1/R$. Given the dependence of the mass on the radiii, we see that the requirement of moduli spaces with finite volume seems to imply the existence of extended objects in the spectrum of the quantum theory.

A finite order identification $R \sim 1/R$ creates a singularity at the fixed point, and 
the standard lore about effective theories with singularities in moduli space is that there are extra states becoming massless at that point. Looking at the mass formula, these new states can't have just KK charges or just winding charges (no matter if they are BPS or non-BPS states). It is conceivable that these new states are non-BPS uncharged states, although so far our experience is that these types of singularities are resolved by BPS states, and I will assume that to be the case. The BPS mass formula dictates then that those are $1/2$ BPS states (the loci where potential $1/4$ BPS states is 
different \cite{Cvetic:1995mx}),  with additional charges, and to vanish at such loci, they should have both $KK$ and winding charge. These states must come in  ${\cal N}=4$ vectormultiplets, since mutliplets with highest spin 3/2 are 1/4 BPS, and  as recalled in \cite{Hull:1995mz},  there are general arguments against multiplets with helicities larger than $3/2$ going massless. 

We have presented some plausability argument (certainly opened to improvement) that there are $1/2$ BPS $W^+$ bosons electrically charged under a lineal combinations of $U(1)_{KK}$ and $U(1)_B$. Since $\Gamma$ is a lattice, it must be a finite index subgroup of $SO(d,d+r;\bZ)$, and by a rotation, there must also be singularities due to massive $W^+$ bosons charged under the $U(1)$s in the Abelian part of the 10d gauge group. Looking at their mass, these become massless if some Wilson line is turned off, so they would have to exist already  in the 10d, but that is impossible for the two Abelian theories, since the gauge group was fixed by anomaly cancellation, so there is no room for enhancing any $U(1)$ in the Abelian factor of the 10d gauge group to $SU(2)$.

\section{The repulson singularity in ${\cal N}=4$ SUGRA}
In this section we again point out a common problem for all four supergravities at hand, that gets repaired by a known non-Abelian mechanism in the heterotic theories, while this cure is not available for the two Abelian theories. While in the previous section we were considering possible singularities in moduli space, now we look at singularities in space-time. We will discuss a type of solution with naked singularities, the repulson singularity, common to all four of these theories. These solutions are charged under a $U(1)$, which by the symmetry of the theories we can take to be a  $U(1)$ in the Cartan subalgebra of the gauge group. 

For the two heterotic theories, these singularities of the classical solution are resolved in the quantum theory (either heterotic strings, or its type II dual formulation), through the beautiful enhan\c con mechanism \cite{Johnson:1999qt}: the quantum theory possesses massive $W^+$ bosons, charged under the $U(1)$ in question, that become massless at a radius bigger than that of the singularity, forming a shell. Inside the shell the metric changes dramatically from the classical solution, it's just flat space, and the singularity is excised \footnote{A somewhat similar mechanism was invoked in 
\cite{Dyson:2003zn} to repair over-rotating black holes, and in \cite{Drukker:2003sc} to repair G\"odel-type solution in string theory}. On the other hand, it is clear that the same mechanism can't be at work for a putative quantum theory of either Abelian supergravity: in the explicit sugra solutions we'll see that a Wilson line goes to zero at the enhan\c {c}on radius, so the $W^+$ boson that would cure the singularity should be present and massless in the 10d theory, but that would violate anomaly cancellation.

Let's very briefly review the solutions with repulson singularities and the enhan\c {c}on mechanism 
referring the reader to the literature for more details. The starting point is 10d ${\cal N}=1$ SUGRA coupled to $d=10$ ${\cal N}=1$ SYM with a gauge group $G$ of rank $r$. Compactification
of these theories on $T^6$ yields a $4d$ sugra theory that at generic points in moduli space has gauge group $U(1)^{r+12}$. The scalars $G_{ij}, B_{ij},A_i^I$, $i=1,\dots 6$ and $I=1,\dots, r$ are parameterized by a matrix ${\cal M}$, whose specific form depends on the basis we are using. If we take the $O(6,6+r)$ invariant metric to be
\begin{equation}
L=\left(\begin{array}{ccc}
0 & 1_6 & 0 \\
1_6 & 0 & 0 \\
0 & 0 & -1_{r} 
\end{array}\right)
\end{equation}
then ${\cal M}$ is
 \begin{equation}
{\cal M}=\left(\begin{array}{ccc}
G^{-1} & G^{-1}D-1_6 & G^{-1}A \\
D^TG^{-1}-1_4 & D^TG^{-1}D & D^T G^{-1} A \\
A^T G^{-1} & A^TG^{-1}D & A^TG^{-1}A+1_r
\end{array}\right)
\end{equation}
where $D=B+G+\frac{1}{2}AA^T$. 
Actually, we will be using a different basis, where the $O(6,6+r)$ metric is diagonal
\begin{equation}
\bar L=\left(\begin{array}{ccc}
1_6 & 0 & 0\\
0 & -1_6 & 0 \\
0 & 0 & -1_{r} 
\end{array}\right)
\end{equation}
which induces a change in the expression of ${\cal M}$. Below, when we write expressions for ${\cal M}$, it will be in this rotated basis. A $\Omega\in O(6,6+r)$ rotation acting as
\begin{equation}
{\cal M}\rightarrow \Omega {\cal M}\Omega^T \hspace{.5cm} A_{\mu}^a \rightarrow
\Omega _{ab} A_\mu^b
\end{equation}
leaves the action invariant, and yields a new solution (generically with
different asymptotic ${\cal M}_\infty$), with the same space-time metric and axiodilaton.

Consider now the following family of $O(6,6+r)$ solutions \cite{Behrndt:1995nn}, written in the diagonal basis. They depend on the choice of two charges $Q_g,Q_v$ and two arbitrary unit vectors: $\vec p$, a 6-vector and $\vec n$, a $(6+r)$-vector. 
\begin{equation}
ds^2=-e^{2U}dt^2+e^{-2Ui}d\vec x^2 \hspace{1cm} e^{4\phi}=1+\frac{4|Q_g|}{r}+\frac{2(Q_g^2-Q_v^2)}{r^2}
\end{equation}
\begin{equation}
{\cal M}=1_{12+r}+4e^{4U}\left(\begin{array}{cc}
 A^2 nn^T& AB np^T\\
 AB pn^T& A^2 pp^T
\end{array}\right)
\end{equation}
\begin{equation}
A(r)=\frac{Q_v}{r} \hspace{1cm}B(r)=\frac{1}{\sqrt{2}}
\end{equation}
These solutions have mass $M=|Q_g|/\sqrt{2}$, and if $Q_v^2>Q_g^2$, a naked singularity at
$r_r=\sqrt{2}(Q_v-Q_g)$, the repulson radius. We choose the unit vectors to be $\vec p=(0,0,0,0,0,1)$ and $\vec n=(0,0,0,0,0,1,0,\dots,0)$, we choose $Q_g=0$,  and leave $Q_v$ arbitrary. Comparing with the general expression for ${\cal M}$ in the diagonal basis, we obtain that this solution has a
single excited scalar field, the radius $g_{66}$ of one the directions of the internal torus,
\begin{equation}
g_{66}(r)=\frac{r-\sqrt{2}Q_v}{r+\sqrt{2}Q_v}
\end{equation}
In string theory this would correspond to a soliton with the charges of a winding mode. Since the solutions we started with have ${\cal M}_\infty=1$, all radii at infinity are self-dual, and the enhan\c {c}on location is $r_e=\infty$. We could readily perform a $SO(6+r)$ rotation on this solution, to arrive at a new one, charged under a $U(1)$ of the 10d gauge group. This gives a solution with a Wilson line for that $U(1)$ turned on (besides the internal radius), and since a $SO(6+r)$ rotation doesn't change the position of the enhan\c {c}on radius, it will still have $r_e=\infty$, i.e. the Wilson line only vanishes at infinity. It is not hard to actually construct solutions charged under a 10d gauge $U(1)$ and with the enhan\c {c}on radius $r_e<\infty$, and we proceed to do it now. First, make a $SO(6,6)$ rotation of the 'winding state' solution, changing only the asymptotic value of the internal excited radius, to an arbitrary value $G_{66}^\infty >1$,
$$
G_{66}(r)=G_{66}^\infty\frac{r-\sqrt{2}Q_v}{r+\sqrt{2}Q_v} 
$$
this doesn't change the direction of the charge vector, so the solution is still charged under the same (KK+'winding') $U(1)$, but now the position of the enhan\c {c}on (i.e. the self-dual radius), determined by $G_{66}(r)=1$, is no longer at $r=\infty$,
$$
r_e=\frac{G_{66}^\infty+1}{G_{66}^\infty-1}r_r > r_r
$$
Now, we make a $SO(6+r)$ rotation, changing under which $U(1)$ the solution is charged, and we do it so for the Abelian supergravities, it is $U(1)$ in the Abelian factor of the 10d gauge group. This rotation changes the moduli fields that are excited, and in the new solution we have ($g_{66}$ denotes the moduli field in the new solution, $G_{66}$ in the old one), 
$$
g_{66}(r)=\frac{4}{2+G_{66}(r)+G_{66}^{-1}(r)}\hspace{1cm}
\frac{a(r)^2}{2}=\frac{G_{66}(r)+G_{66}^{-1}(r)-2}{G_{66}(r)+G_{66}^{-1}(r)+2}
$$
We are almost there: this solution is charged under a $U(1)$ in the 10d gauge group, and has a non-zero Wilson line $a(r)$. This Wilson line vanishes when $G_{66}(r)=1$, i.e. at $r_e$. This is also the 
enhan\c {c}on radius for this solution. The only concern is that this solution, at the radius where the Wilson line vanishes, also has the sixth radius self-dual, so we apply a final $SO(6,6)$ transformation that changes the asymptotic radius of this last solution, 
$$
g_{66}(r)=g_{66}^{\infty}\frac{4}{2+G_{66}(r)+G_{66}^{-1}(r)}\hspace{1cm}
\frac{a(r)^2}{2}=g_{66}^{\infty}\frac{G_{66}(r)+G_{66}^{-1}(r)-2}{G_{66}(r)+G_{66}^{-1}(r)+2}
$$
This is the solution we were after. It is charged under a $U(1)$ in the 10d gauge group. It has a naked singularity at $r_r$, and a Wilson line vanishing at a bigger radius, $r_e>r_r$. In the examples that appear in string theory, at $r_e>r_r$ extra massless particles appear, enhancing the U(1) gauge group 
to $SU(2)$ (there are massless $W^\pm$s and monopoles). Now, the candidate states to go massless at this radius $r_e$ are $W^\pm$ bosons, enhancing the $U(1)$ to an $SU(2)$. Since these potential $W^\pm$ bosons are in general massive because a non-vanishing Wilson line, they should be present in the 10d theory. This is fine for the two heterotic SUGRA theories, since the 10d gauge groups are non-Abelian, and we know that at specific points of the moduli space, that non-Abelian character is recovered. On the other hand, ${\cal N}=1$ $d=10$ sugra coupled to $U(1)^{496}$ 
and $E_8\times U(1)^{248}$ can't afford any of its $U(1)$ factors enhanced to $SU(2)$, since that would violate the constraints imposed by anomaly cancellation.

A word of caution. For the two Abelian theories, we can't perform the thought experiment of creating these solutions by bringing shells of matter from infinity, since {\it a priori} we don't know if the quantum theories have charged matter under the appropriate $U(1)$ (in fact, we want to argue that they don't !). So in these theories, the solutions with naked singularities exist forever, but it might not be possible to even try to create them dynamically.

As for trying to fix the singularity with other mechanism, using just supergravity \cite{Johnson:2001wm},  one can construct supergravity solutions that have the same asymptotics, but excise a region that contains the naked singularity. To do so consistently, Israel matching
conditions dictate what stress energy tensor we should put at the junction. There are also matching conditions for the scalar and gauge fields. In particular, the matching condition for the gauge field requires that the shell is made of matter charged under the $U(1)$ of the solution. Furthermore, one can try to play with more elaborate excisions: other types of interior metric, a number of concentric shells...,  but since in the original case the tension of the brane became null at the enhan\c {c}on radius, it seems plausible that other excisions using patches with positive energy density somewhere will require negative density energy elsewhere, so it might be hard to find another solution whose stress-energy tensor at the junction or interior does not violate any energy condition\footnote {I would like to thank S. Ross for this observation.}.

\section{Acknowledgements} 
I would like to thank Asad Naqvi for conversations that originated this work. I would also like to thank Greg Moore, Dave Morris, Amanda Peet, Simon Ross, Erik Verlinde and Cumrun Vafa for comments and discussions. This work was supported in part by by grants FPA2007-66665C02-02 and  DURSI 2005-SGR-00082 and by a Ram{\'o}n y Cajal fellowship.

\end{document}